# CONTAMINATION BY METALLIC ELEMENTS RELEASED FROM JOINT PROSTHESES

*E. CHASSOT[1], J.L. IRIGARAY[1], S. TERVER[2], G. VANNEUVILLE[3]*

*[1] Laboratoire de Physique Corpusculaire de Clermont-Ferrand, CNRS – IN2P3, Université Blaise Pascal, F-63177 Aubière Cedex – France*

*[2] CHRU, Service Orthopédie, Université d'Auvergne, F-63100 Clermont-Ferrand, France*

*[3] Laboratoire d'anatomie, Université d'Auvergne, F-63100 Clermont-Ferrand, France*

Corresponding author: E.Chassot

Laboratoire de Physique Corpusculaire de Clermont-Ferrand, CNRS – IN2P3, Université Blaise Pascal, F-63177 Aubière Cedex – France

Tel : + 33 (4) 73 40 76 51

Fax : + 33 (4) 73 26 45 98

Email: chassot@clermont.in2p3.fr

**Abstract**

When a metallic implant is in contact with human tissues, the organism reacts and a corrosion process starts. Consequently, we might observe liberation of metallic debris and wear. Our purpose is to measure the contamination and the migration of these metallic elements in the surrounding tissues of the implant.

Two types of samples have been studied. First type is sample taken on post-mortem tissues around prostheses to study contamination gradients. Second type is sample taken on pathologic joints on periprosthetic capsular tissues in surgical conditions. These allow estimating contamination degree.

The experiments were made on a Van de Graaff accelerator located at CERI (*Centre d'Etude et de Recherche par Irradiation, Orléans, France*). We measure elemental concentrations resulting from the contamination of the surface of each sample. Results are analysed in function of the pathology and the type of implants.

According to the pathology and the location of the sampling, these measurements show a very heterogeneous contamination by metallic elements under particles and/or ionic species which can migrate through soft tissues by various mechanisms.

Key words: Hip prosthesis, metallic release, ionic release, PIXE

## Introduction

In orthopaedic surgery, metallic implants, like total hip prostheses or total knee prostheses, are routinely used in articular diseases. A few years after their implantations, some prostheses inserted for joint disease may show adverse reactions such as: periprosthetic osteolysis, instability of the fixation or fracture. These metallic prostheses may undergo degradations. The influence of metallic ions released on periprosthetic tissues shows a significative issue. Many measurements were carried out on blood and serum; however the contamination of periprosthetic tissues is not well understood. A qualitative and quantitative analysis is necessary to determine the degree of contamination in soft tissues.

In this paper, we study contamination induced by metallic element released by joint prostheses. The process of analysis of biological tissues requires the examination of different parts. We analyse some tissues near femur, in contact with metallic prostheses, and measure by sensitive physical methods the importance of metallic release (1). A first experiment by neutron activation analysis (NAA) demonstrates the contamination of adjacent tissues surrounding implants (2, 3, 1, 4, 5, and 6). Particles Induced X-ray Emission (PIXE) has been used to analyse major, minor and trace elements contained in human tissues. This method evaluates the micrometer level metallic contamination introduced by the prosthesis (7, 8, 9, and 10).

## Materials and methods

### Prostheses analysis by Atomic Emission Spectroscopy

To identify the type of implanted prosthesis, we have used spark spectrometry. The principle of this technique consists in exciting an atom. When this atom returns to the fundamental state, it releases energy as luminous radiation where the wave number characterises the respected atom. A high energy given by a spark under argon with a 15-30 kV voltage volatilises a few quantity of the material. The excitation is the result of collisions in the plasma (11).

### Samples preparation

We analyse two types of samples: muscular and capsular tissues near the femoral head.

The first one is taken from anatomical subjects (post-mortem samples noted PMS). Observations do not show coloration induced by the presence of metallic elements neither abnormal macroscopic movements of the prosthesis within the bone. We postulate that there are no adverse reactions. We analyse samples from per prosthetic tissues of 5 cases:

- PMS1 is a specimen removed near a prosthesis of hip cemented out of titanium alloy,
- PMS2 is a tissue sample removed near a prosthesis of hip, aiming, cemented with a polyethylene cup covered with titanium,
- PMS3 and PMS4 result from the same patient but on the right and left hip. PMS3 is cemented cobalt alloy prosthesis. PMS4 is also out of cobalt alloy but taken near the prostheses titanium pins.
- PMS5 is a cemented prosthesis out of stainless steel with a polyethylene cup. The tissues in the vicinity are black coloured.

In these cases, the samples are taken all around the implant at different place locate with the number 1 to 6 on Fig.1.

The second type of sample is taken during surgical procedures indicated at the time of implant loosening or inflammatory reactions. There are 4 different prostheses (surgical samples noted SS):

- SS1 is a capsular tissue of hip removed near an intermediate prosthesis cemented out of stainless steel with a polyethylene cup. The prosthesis was unstable, the tissues were black coloured and there was osseous lysis.
- SS2 is a capsule removed around a cemented prosthesis cobalt alloy hip. As previously, the prosthesis is unstable, the tissues are black coloured and there is the presence of an osseous lyses,
- SS3 and SS4 are two prostheses of titanium alloy hip. SS3 is a cemented prosthesis with a ceramic head. At its proximity, there are an osseous lyses and a lesion on the surface stem. SS4 is a prosthesis concerned unstable covered with alumina.

For these cases, the type of implant, the insertion period and all symptoms which occurs the procedure of operation (inflammatory reaction, loosening of the implant, instability…) are noted.

A control sample (SS5) is taken from capsular tissue in a patient at the first stage of primary hip prosthesis procedure.

All the tissues are taken off with a stainless steel scalpel, which appears as being the less polluting tool (5). The samples are wrapped in a plastic bag and keep in a freezer, waiting for analyses.

Before analysis, each sample is freeze-dried between two-polyethylene plates in order to have a planar surface. This condition is necessary to allow irradiation. In all cases, the thicknesses of the sample are 2 mm. They were coated by a thin carbon layer to reduce charge effects at the surface of the sample during irradiation.

### PIXE method

This technique is based on the X-ray spectrometry produced by a charged particle beam that irradiates a target (12). This method presents some advantages comparing to another method like electron fluorescence:

- the sensitivity is around µg/g (depending of the matrix),
- it is a multielementary technique,
- the dimension of the beam is adjustable.

The quantitative analysis by PIXE microprobe consists in measuring the intensity of the characteristic X-rays of the elements in the target and its conversion in concentration.

We have used GUPIX software developed by Campbell (13) to determine the concentration of metallic elements in the sample. Peaks at a given energy characterise an element.

### Experimental device

Van de Graaff accelerator (VDG) is located at CERI *(Centre d'Etude et de Recherche par Irradiation, Orléans, France*). This VDG can accelerate protons (Fig. 2) with an energy beam of 3 MeV and an intensity of 2 nA. The beam diameter is about 1 mm. We make several measurements on the target surface to take into account tissue heterogeneity. The detection of X-rays is made with Si (Li) detector placed at 135° from the direction of the incident beam. In front of this detector, we place a carbon filter, which has the function to reduce the intensity of the light elements, dead time and allows trace element detection.

### Analysis methods

In order to quantify the contamination, the weighted mean concentration (µg/g) of major elements released from the prosthesis is calculated for each sample. We make 9 measurements on a given area to have a better representation of the contamination. In table 1 and 2, < $C_{moy}$ > represents the weighted average concentration for all the measured points made around the implant. Max ($C_{moy}$) is the maximum concentration obtained on the surface in a special location around the implant. Max (C) is the maximum concentration of a measured point that we can found around the implant. Most of time, this point is at the same location that Max ($C_{moy}$).

The results show a great heterogeneity of measurements which indicates that some points are far from the mean. We tried to separate this phenomenon in two categories. 1) The mean concentration can represent the global contamination. 2) The maximum concentration can be associated with the presence of metallic fragments. The analysis is separated in two points: the first one concerns post-mortem samples and the second one surgical sample.

## I: Post-mortem tissue

### Results

The analyses of post-mortem tissues near implant (PMS) show the presence of metallic elements in adjacent tissues.

In table 3, the proportion of concentration above the limit of detection is presented. The results lie between 17% and 70% over the limit of detection. The repartition of the metallic elements around the prosthesis in post-mortem tissues shows that the largest migration takes place around the head of the prostheses in the trochanteric insertion of muscle (figure 1).

In the Fig 3, we have represented the 9 measured points for PMS5. The graph represents the variation of the contamination on a 9 mm² area. Each point has 1 mm² area and all the points are adjacent the one with the other. We have represented Cr, Fe, Ni, Mn, Co and Mo which are the major elements of the prostheses. The surface analysis shows high variability in concentration between these points. The highest concentration are at the same location for all the tested elements (for instance the coordinates: 3-3; 3-2; 2-3), but the cobalt shows high level at another location (for instance 1-3) which doesn’t correspond to major elements present in the implant (Cr-Fe-Ni).

In table 1, straightforward results show that on PMS the only significant mean concentration is found for iron. But the observation of the maxima shows the existence of contamination by metallic elements in particular location.

**Discussion**

In most of cases, post-mortem samples present no inflammatory reactions to the implant. The mean contamination all around the implant is not significant. But, if we consider a particular location, the contamination is more important.

The comparison between the maximum concentration (Max ($C_{moy}$)) and the maximum point (Max (C)) shows that the migration and the contamination are not regular along the implant. The evolution of the contamination depends on the location of the sample. In most cases, the major concentration is observed near the head of the prostheses in the trochanteric insertion of muscle. In this case, the distance between the tissue and the implant is not so far. This is not the case for the sample along the stem where the elements must cross cement and cortical bone. But sometimes, high contamination may be seen along the stem: we postulate that it is an effect of bone circulation.

The studied stainless steel prosthesis (PMS5) shows correlation between chromium, iron and nickel (Fig 3). The proportion of these metals compared to those found in prostheses (table 4) gives information about possible correlations between different elements. So we can conclude that we are in presence of fragments or cluster because of the ratios which are much closed to them of the prosthesis.

**II: Surgical samples**

**Results**

The study of tissue of reference carries out under the same conditions as for pathological fabrics. The metallic elements present are iron and zinc. Other metals are below the limit of detection of our method (<20 µg/g).

Table 2 shows concentrations obtained on surgical samples. The main constitutive elements of the implants are found in surrounding tissues. Special mention must be made for chromium in SS1 and SS2 and cobalt in SS2. Although titanium is found around SS1 (Stainless steel). Iron is always at a high level whatever the implant is, maybe due to blood.

Table 3 shows the proportion of concentration above the limit of detection and the results show that four of the five surgical samples have element concentration over 70%. The difference of percentage varies according to the pathology and the state of the sample (black coloured or not).

In table 5, comparison of major metallic ratio between prosthesis and surgical tissue has been made and show a good correlation for chromium-iron and nickel-iron, but less good correlation when chromium and nickel are concerned.

## Discussion

Results show that the contamination of adjacent tissues depends on the implant. The mean concentrations are not very high comparing to the maximum. These averages mask points of high concentration (table 1 and 2). By comparison of metallic ratio, there is good agreement between Cr/Fe and Ni/Fe. Concerning the Cr/Ni, the ratio is upper in tissue to the prosthesis. This can be due to the fact that there is a high chromium concentration in tissue or less concentration in nickel. It can be a consequence of the evolution of these elements in organism, if they are stocked by cells or if they are transfer towards other organs to be eliminated.

We observe that the contamination is not regular and depends on the type of implant and the state of the tissue. For surgical samples, the patient had a reaction to the prostheses. The presence of metallic elements is important and may correspond to a contamination due to the corrosion of the tissue. If the contamination is homogeneous on a surface and if the quantity of element doesn't appear as important, it might correspond to an ionic contamination. Higher concentration may point out another possible type of contaminations resulting from the release of wear metallic fragment. Whatever the case, the majority of the contamination is above the limit of detection (Table 1, Table 2).

## Conclusion

The corrosion has already been studied by Betts and Shalgaldi (14, 15). Effects of metallic elements released by prostheses on biological behaviour was described by Allen (16) as well as the contamination in other tissues was observed by Henning, Schnabel & Urban (17, 6, 18) and contamination of body fluids by Liu (19). In this work, we have measured the contamination in the adjacent tissue induced by metallic implants and we confirm the contamination by metallic elements in periprosthetic tissues.

The study of the control sample, where there is no metallic implant, indicates that the method used to take and to conserve the samples does not induce contamination (5). Furthermore, preliminary trials show that the irradiation times have no effect on the concentration.

To show different types of contamination, an analytical method has been used (20). The weighted average concentration permits to evaluate the quantity of metallic elements contained in tissues. The separation in two categories of measurements investigates the heterogeneity of repartition. In the case of post-mortem samples, the major contamination seems to come from ionic species. Concerning surgical tissues, the metallic element concentration is higher and migration of wear debris is presumed. This can explain that post-mortem tissues concern patient who had no prosthesis problem, no inflammatory reaction even when in surgery, the patient can have a reaction against the implant. The elements released from the prostheses might be eliminated from the organism by cells. Sites with high concentrations can correspond to accumulation of the elements or can result from an exterior contribution. On the other hand, in surgical samples a higher correlation between elements is observed. It demonstrates the possible presence of fragments.

The differences between the mean concentration (< $C_{moy}$ >) and the maximum (Max ($C_{moy}$)) are smaller in surgical samples than in post-mortem tissues. This result comes from the heterogeneity and points of high concentration.

If we look for correlation between elements, in most of the cases, all the metals are present at the same location and the metallic ratio is nearly the same than in the prosthesis. The presence of these elements has already been studied by SEM and TEM methods (21, 22). In this study, we were looking for the quantification of these phenomenons with sensitive physical methods. The distribution of these metallic elements on a surface near the implant has been investigated. However, the evolution of the metal released through the muscle to other tissues should be analysed to give more information about the migration. Complementary studies are needed to conclude on contamination nature and depth. We are looking for correlation between major elements that could detect prosthesis fragments.

We acknowledge G.Blondiaux, director of the CERI at Orléans (France) and all persons who help us to realise the experiment, mainly T. Sauvage, Y. Tessier and O. Wendling.

Table 1: Contamination of post-mortem samples near an implant. The concentrations are given in µg/g. (LD: Limit of Detection)

| | | PMS1<br>Ti | PMS2<br>Ti + Co | PMS3<br>Co-Cr | PMS4<br>Co-Cr | PMS5<br>Cr-Fe-Ni |
|---|---|---|---|---|---|---|
| **Ti** | **< $C_{moy}$ >** | **32 ± 7** | **30 ± 10** | **26 ± 6** | **<LD** | **<LD** |
| | Max ($C_{moy}$) | 80 ± 14 | 703 ± 21 | 73 ± 18 | 28 ± 16 | 24 ± 23 |
| | *Max (C)* | *607* | *17741* | *1887* | *824* | *187* |
| **Cr** | **< $C_{moy}$ >** | **<LD** | **<LD** | **<LD** | **<LD** | **51 ± 12** |
| | Max ($C_{moy}$) | < LD | <LD | 28 ± 24 | 225 ± 18 | 4816 ± 39 |
| | *Max (C)* | *26* | *198* | *172* | *865* | *59549* |
| **Fe** | **< $C_{moy}$ >** | **116 ± 5** | **376 ± 49** | **129 ± 14** | **52 ± 3** | **220 ±18** |
| | Max ($C_{moy}$) | 168 ± 20 | 957 ± 41 | 305 ± 23 | 481 ± 22 | 13409 ± 71 |
| | *Max (C)* | *305* | *4079* | *589* | *2292* | *56898* |
| **Co** | **< $C_{moy}$ >** | **<LD** | **<LD** | **<LD** | **<LD** | **<LD** |
| | Max ($C_{moy}$) | <LD | <LD | <LD | 474 ± 31 | <LD |
| | *Max (C)* | *<LD* | *<LD* | *26* | *705* | *344* |
| **Ni** | **< $C_{moy}$ >** | **13 ± 2** | **<LD** | **<LD** | **<LD** | **<LD** |
| | Max ($C_{moy}$) | 25 ± 9 | <LD | <LD | 15 ± 12 | 1873 ± 22 |
| | *Max (C)* | *31* | *<LD* | *163* | *56* | *696* |

Table 2: Contamination in capsular surgical samples. The concentrations are given in µg/g. (LD: Limit of Detection)

| | | **SS1**<br>Cr-Fe-Ni | **SS2**<br>Co-Cr | **SS3**<br>Ti | **SS4**<br>Ti | **SS5**<br>Witness |
|---|---|---|---|---|---|---|
| **Ti** | **$< C_{moy} >$** | **561 ± 53** | **<LD** | **322 ± 38** | **763 ± 90** | _ |
| | *Max (C)* | *1057* | *13* | *368* | *1734* | _ |
| **Cr** | **$< C_{moy} >$** | **1614 ± 349** | **3523±1026** | **<LD** | **357 ± 163** | _ |
| | *Max (C)* | *2672* | *4856* | *<LD* | *1144* | _ |
| **Fe** | **$< C_{moy} >$** | **5010 ± 833** | **905 ± 178** | **492 ± 167** | **4461 ± 600** | **93 ± 31** |
| | *Max (C)* | *7778* | *1196* | *969* | *6303* | *129* |
| **Co** | **$< C_{moy} >$** | **117 ± 24** | **1453 ± 214** | **<LD** | **57 ± 17** | _ |
| | *Max (C)* | *134* | *2284* | *<LD* | *109* | _ |
| **Ni** | **$< C_{moy} >$** | **818 ± 132** | **60± 31** | **7 ± 4** | **247 ± 74** | _ |
| | *Max (C)* | *1277* | *116* | *23* | *432* | _ |

Table 3: Concentration fraction over the limit of detection (LD) for metallic elements (Ti, Cr, Co, Fe, Ni, Mo)

| **PMS1** | **PMS2** | **PMS3** | **PMS4** | **PMS5** |
|---|---|---|---|---|
| 60% | 70% | 18% | 17% | 70% |
| **SS1** | **SS2** | **SS3** | **SS4** | **SS5** |
| 100 % | 100% | 69% | 76% | 18% |

Table 4: Comparison between metallic ratios in prosthesis and in post-mortem sample

| | Cr/Ni | Cr/Fe | Ni/Fe |
|---|---|---|---|
| Prosthesis | 1.4 ± 0.1 | 0.3 ± 0.1 | 0.2 ± 0.1 |
| Tissue | 3.1 ± 0.3 | 0.5 ± 0.1 | 0.2± 0.1 |

Table 5: Comparison between metallic ratios in prosthesis and in surgical tissue

| | Cr/Ni | Cr/Fe | Ni/Fe |
|---|---|---|---|
| Prosthesis | 1.37 ± 0.01 | 0.27 ± 0.01 | 0.19 ± 0.01 |
| Tissue | 2.09 ± 0.14 | 0.34 ± 0.02 | 0.16 ± 0.01 |

Figure 1: Location of the sampling made near a metallic hip prosthesis. Numbers correspond to sample identification.

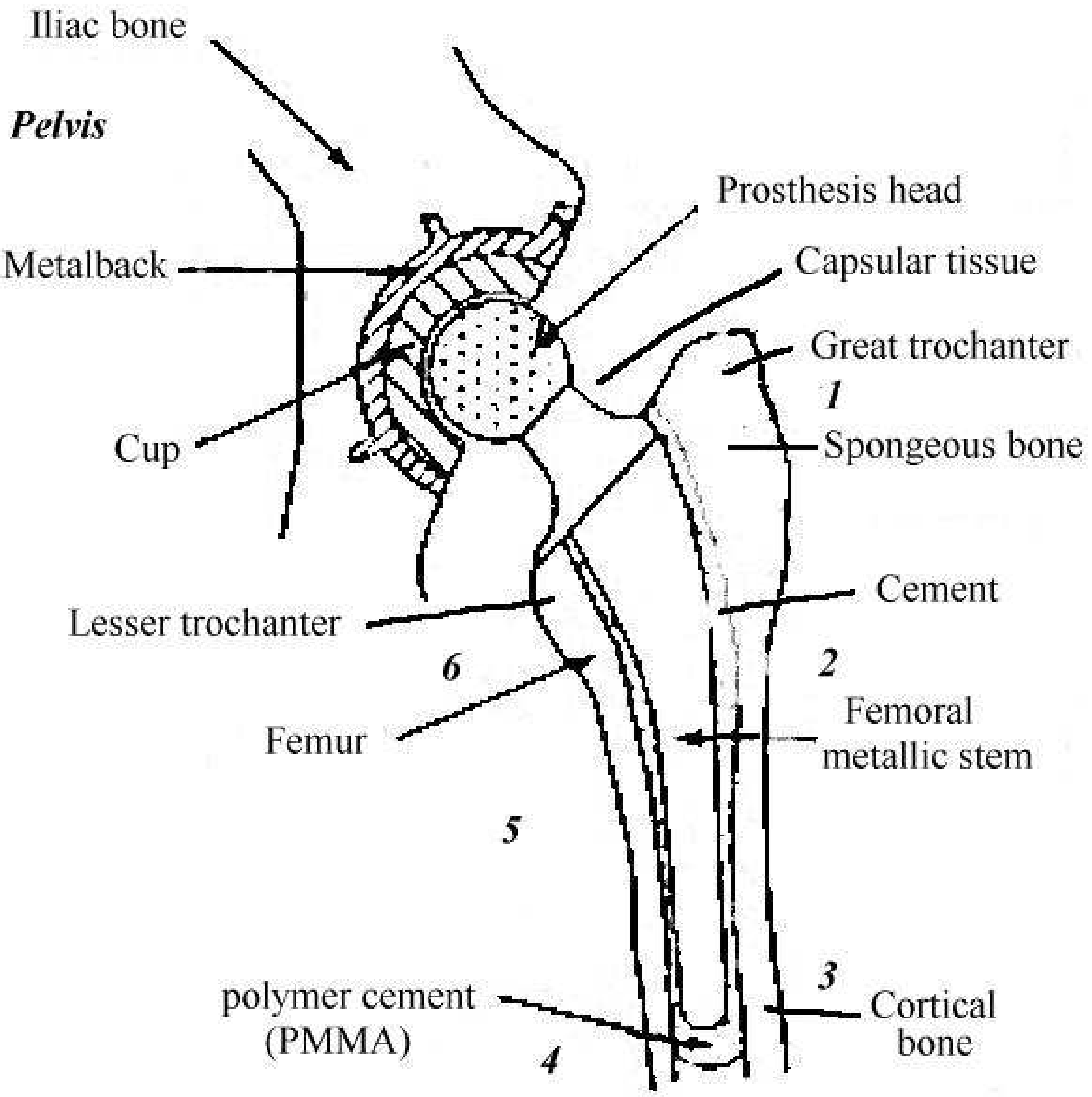

Figure 2: Van de Graaff accelerator

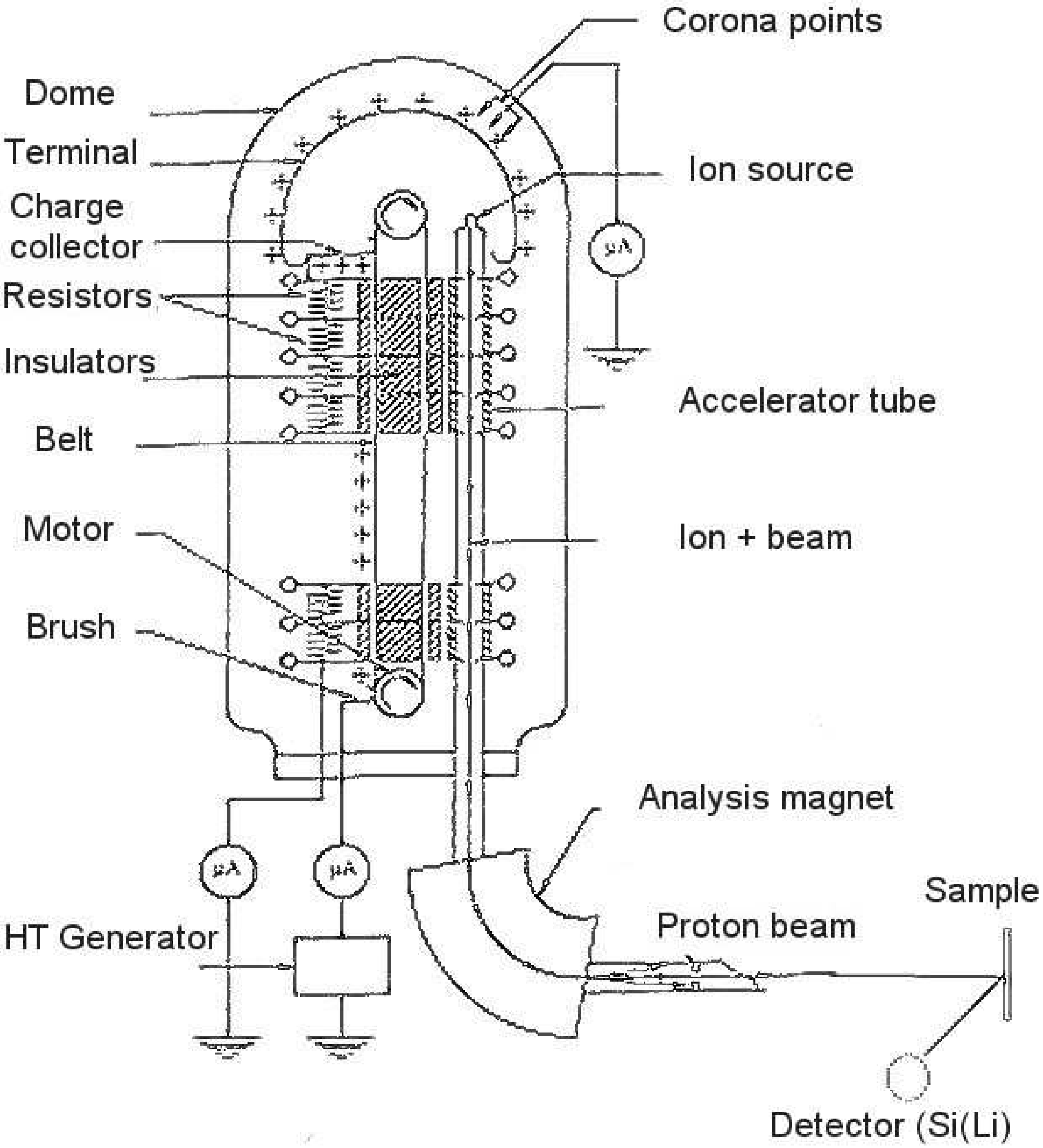

Figure 3: Elemental distributions on PMS5 tissue surface.

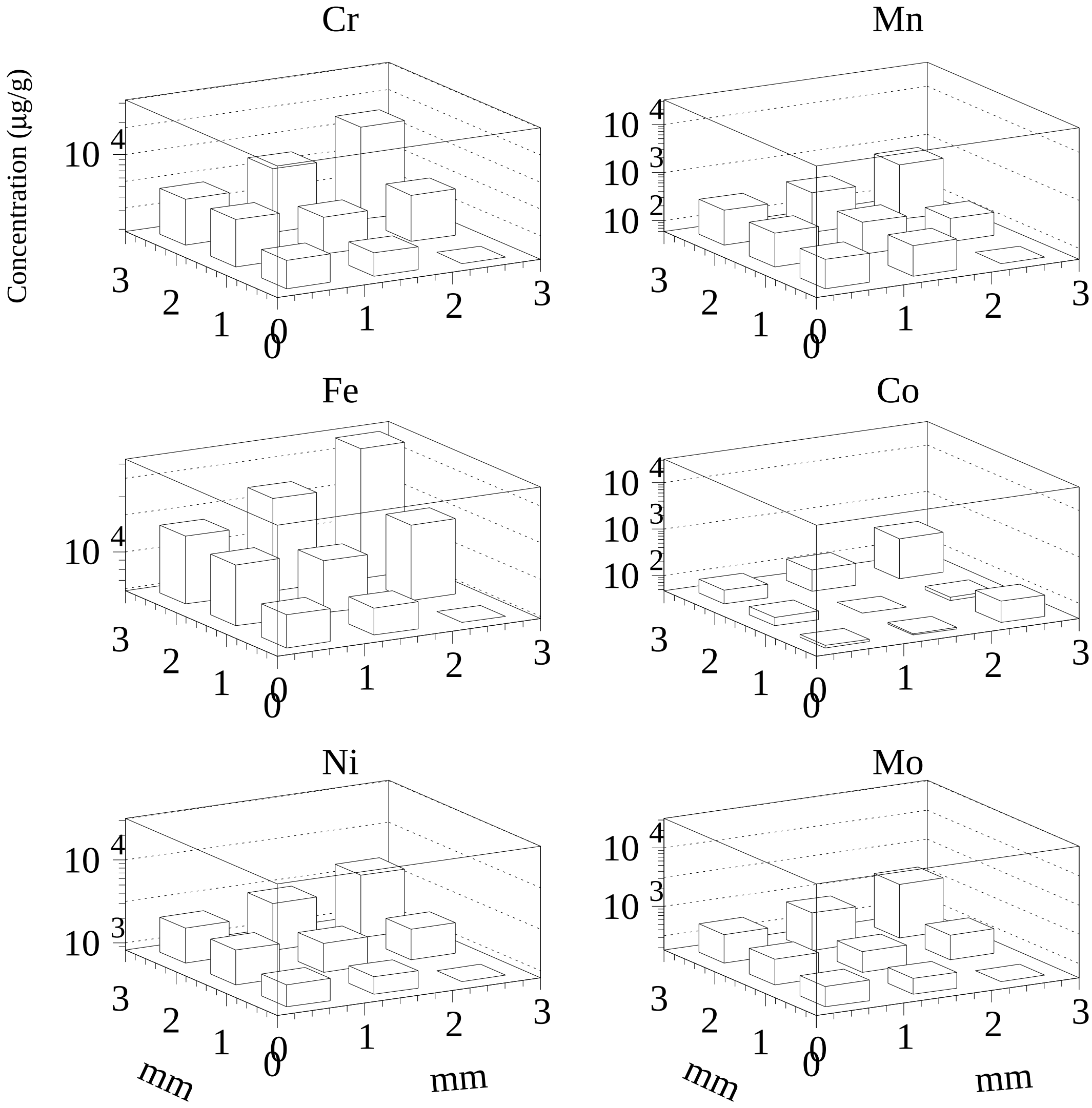

Table 1: Contamination of post-mortem samples near an implant. The concentrations are given in µg/g. (LD: Limit of Detection)

Table 2: Contamination of capsular surgical samples. The concentrations are given in µg/g. (LD: Limit of Detection)

Table 3: Concentration fraction over the limit of detection (LD) for metallic elements (Ti, Cr, Co, Fe, Ni, Mo)

Table 4: Comparison between metallic ratios in prosthesis and in post-mortem sample.

Table 5: Comparison between metallic ratios in prosthesis and in surgical tissue

Figure 1: Location of the sampling made near a metallic hip prosthesis. Numbers correspond to sample identification.

Figure 2: Van de Graaff accelerator

Figure 3: Elemental distributions on PMS5 tissue surface.